\documentclass{ws-ijqi}
\usepackage{amsmath} 
\usepackage{graphicx}
\newcommand{\Aptilde}{\tilde{{\cal A}}_p}				% 
\newcommand{\Ap}{{\cal A}_p}	
\newcommand{\psitilde}{\tilde{\psi}}				% biphoton amplitude in the Fourier domain

\newcommand{\rr}{\vec{r}}
\newcommand{\q}{\vec{q}}
\newcommand{\w}{\vec{w}}
\newcommand{\wprime}{\vec{w}\,^{\prime}}
\newcommand{\vxi}{\vec{\xi}}

\newcommand{\be}{\begin{equation}}
\newcommand{\ee}{\end{equation}}
\newcommand{\bea}{\begin{eqnarray}}
\newcommand{\eea}{\end{eqnarray}}
\newcommand{\bse}{\begin{subequations}}
\newcommand{\ese}{\end{subequations}}
\newcommand{\A} {\hat{A}}
\newcommand{\e}{\mathrm{e}}

\newcommand{\Om}{\Omega}

\newcommand{\sinc}{\mathrm{sinc}}

\newcommand{\Depw}{\Delta_{\rm pw}}

\begin{document}
\markboth{Gatti et al.}
{Spatio-temporal entanglement of twin photons: an intuitive picture}
%%%%%%%%%%%%%%%%%%%%% Publisher's Area please ignore %%%%%%%%%%%%%%
%\catchline{}{}{}{}{}
%%%%%%%%%%%%%%%%%%%%%%%%%%%%%%%%%%%%%%%%%%%%%%%%%%%%%%%%%%%%%%%%%%%
%
\title{Spatio-temporal entanglement of twin photons: an intuitive picture}
\author{Alessandra Gatti}

\address{Istituto di Fotonica e Nanotecnologie, Consiglio nazionale delle Ricerche, Piazza Leonardo da Vinci 32, Milano, Italy
\\ Alessandra.Gatti@mi.infn.it}

\author{Lucia Caspani\footnote{Present affiliation:INRS-EMT, 1650 Boulevard Lionel-Boulet, Varennes, Quebec J3X 1S2, Canada
}, Tommaso Corti, Enrico Brambilla }

\address{Dipartimento di Scienze e Alta Tecnologia, Universit\'a  dell'Insubria, Via Valleggio 11, Como, Italy}

\author{Ottavia Jedrkiewicz}

\address{Istituto di Fotonica e Nanotecnologie, Consiglio nazionale delle Ricerche, Piazza Leonardo da Vinci 32, Milano, Italy}

\maketitle

%\begin{history}
%\received{Day Month Year}
%\revised{Day Month Year}
%\accepted{Day Month Year}
%\comby{(xxxxxxxxxx)}
%\end{history}

\begin{abstract}
We draw an intuitive picture of the spatio-temporal properties of the entangled state of twin photons, where they are described as classical wave-packets. This picture predicts a precise relation between their temporal and transverse spatial separations at the crystal output. The  space-time coupling described by classical arguments turns out to determine in a precise way the spatio-temporal structure of the quantum entanglement, analysed by means of the biphotonic correlation and of the Schmidt dimensionality of the entanglement. 
\end{abstract}


\section{Introduction}	
The process of parametric down-conversion (PDC) occurring in a nonlinear crystal is a widely employed source of entangled photons. One of the appealing feature of this process is the possibility of generating  {\em high-dimensional entanglement} ,  both  because  various degrees of freedom of the photon pair are entangled (polarization, time-energy, position-momentum), and because spatial and temporal entanglement is realized in a high-dimensional Hilbert space, due to the naturally ultra-broad bandwidths of PDC.   
\par
Depending on the application, traditional approaches often focus on a single degree of freedom at a time,  or even when considering them simultaneously, treat them as independent. However, as for many  nonlinear optical processes, PDC is ruled by phase matching, which establishes an angular dispersion relation that links the frequencies and the angles of emission of the generated photons in a non-factorable way. A non-trivial consequence is that, unless the spectral bandwidths considered are very narrow, it is not possible in general  to  speak about a purely temporal or spatial entanglement of the state, but rather the entanglement  involves the spatial and temporal degrees of feedom of twin photons in a non separable way. Recent theoretical \cite{gatti2009,caspani2010,brambilla2010,brambilla2012,gatti2012} and experimental \cite{jedr2012a,jedr2012b} works have indeed demonstrated that the correlation function of twin photons or twin beams generated in collinear phase matching conditions have a peculiar X-shape in the spatio-temporal domain,  which entails the attracting  possibility of controlling  the temporal properties of entanglement by acting on the spatial degrees of freedom of twin photons, or viceversa. 
\par
In this work we shall introduce a somehow intuitive picture of the origin of such spatio-temporal entanglement, where twin photons are pictured as quasi-classical wave-packets. We shall demonstrate how the phase matching mechanism, i.e. the conservation of energy and  momentum in the microscopic process, involves a precise relation between the temporal delays and the transverse spatial separation of the  members of  a pair, which turns out into a characteristic geometry of their quantum  correlation in the space-time domain. 
\par
From a more formal perspective, the second part of this work will address the issue of the  degree of the entanglement of the state  
quantified  by the so-called Schmidt number.  A comparison with the results obtained  in either a purely spatial or a purely temporal model shows that the degree of entanglement of the PDC state in the full 3D spatio-temporal domain cannot in general be trivially reduced to the product of the Schmidt numbers obtained in models with lower dimensionality, unless the detected bandwidth is very narrow \cite{gatti2012}.  Interesting differences between the case of collinear and non-collinear phase matching will be analysed, giving also in this case evidence of the role of phase matching in determining the geometrical properties of the state. 

\section{Spatio-temporal description of parametric down-conversion}
A spatio-temporal model for parametric down-conversion has been described in detail in \cite{gatti2003,caspani2010}. Although some of the following results can be generalized to the high gain regime of PDC, and to type II phase matching,  for the sake of simplicity we shall restrict to type I PDC in the  ultra-low gain regime, where very few photon pairs with identical polarization are generated in each spatio-temporal mode.

We consider an intense laser pump field of central frequency $\omega_p$, propagating along the z-axis into a slab of $\chi^{(2)}$ nonlinear crystal of length $l_c$, cut for type I phase matching.   The model \cite{caspani2010} is formulated in terms of a propagation equation along the crystal slab for the quantum field operator $\A (\w,z)$,  associated with the down-converted field of central frequency $\omega_p/2$. The variable 
\be
\w = (\q,\Om) 
\ee
is a short-hand notation for the 3D spatio-temporal {\em Fourier}  coordinate, where $\q=(q_x, q_y)$ is the photonic wave-vector in the direction transverse to the propagation axis $z$, and $\Om= \omega- \omega_p/2$ is the frequency shift from the central frequency. 
In the following we shall denote the 3D coordinate in the {\em direct}  domain as 
\be
\vxi= (\rr, t)
\ee
where $t$ is time and $\rr= (x,y)$ is the spatial coordinate spanning the transverse plane at the crystal exit face (near-field plane), with the convention
$\w \cdot \vxi= \q \cdot \rr  - \omega t$. 

In the undepleted pump approximation, where  the pump can be considered a c-number field, the propagation equation can be formally solved  as a perturbative expansion \cite{caspani2010} in powers of the  
dimensionless coupling strength $g=\chi^{(2)} \alpha_{p}l_{c}$ , where $\chi^{(2)}$ is a parameter  proportional to the  second-order susceptibility of the medium and $ \alpha_{p}$ is the peak value of the pump amplitude. 
In the ultra-low gain regime, defined by  $g \ll 1$,   the first order of such expansion provides the following  input-output relation linking the signal field operator at the output face of the crystal with the one at the input face \cite{caspani2010}: 
\bea 
	\A(\w, l_c)&=& \e^{ik_{sz}(\w)l_{c}} 
	\left[ \A(\w,0) \right.  \nonumber \\
&+& \left. g \int \frac{d^3 \w}{(2\pi)^{3/2}} \Aptilde(\w+\wprime,0) \e^{-i\frac{\Delta(\w,\wprime)}{2} } \sinc \frac{\Delta(\w,\wprime) }{2}  \A^{\dagger}(\wprime, 0) \right]\, .
\label{inout}
\eea
In this equation $\Aptilde$ is the Fourier transform of the pump beam profile at the crystal input face: 
$
\Aptilde (\w) := \int  \frac{ d \vxi} { (2 \pi)^{3/2} }   \Ap (\vxi,0)    e^{-i \vxi\cdot \w} \: ,
$
where normalization is such that $\Ap (\vxi =0, z=0) =1$; $\Delta$ is the phase mismatch function, which accounts for the conservation of longitudinal momentum in the microscopic PDC process 
\be
\Delta (\w,\wprime) = \left[ k_{sz} (\w) + k_{sz} (\wprime) - k_{pz} (\w +\wprime)\right] \, l_c
\label{Delta}
\ee
$k_{sz}$ being the longitudinal component of the  signal wave vector, $k_{pz}$ the analogous quantity for the pump. This quantity
determines the efficiency of each elementary down-conversion process, in which a pump photon in the mode  
$\w +\wprime=(\q+ \q\,^\prime, \Om+\Om^\prime)$ , splits into two twin photons in the modes  $w=(\q,\Om)$ and $\wprime=(\q\,^\prime,\Om^\prime)$,respectively,  with conservation of the energy and transverse momentum.   

From the relation \eqref{inout}, assuming that the input signal field is in the vacuum state, the probability amplitude of generating a pair of photons in the Fourier spatio-temporal modes $\w$ and $\wprime$, i.e. the {\em biphoton amplitude in the Fourier domain},  is readily calculated
\bea
\psitilde(\w,\wprime) &:=& \langle \A(\w,l_c) \A(\wprime,l_c) \rangle  \label{psitilde1}\\
&=& \frac{ g} { (2 \pi)^{3/2} }   \Aptilde(\w+\wprime,l_c) \e^{ i\frac{\Delta(\w,\wprime)}{2} } \sinc \frac{\Delta(\w,\wprime) }{2}  
\label{psitilde}
\eea
where the expectation value in \eqref{psitilde1}  is taken over the input vacuum state and 
$\Aptilde(\w,l_c) = \e^{ i k_{pz} (\w) l_c} \Aptilde (\w,0) $ is the profile of the pump field after it has propagated linearly to the crystal end face. 

So far we  used a field formalism, where the field operators evolve from the input to the output face of the crystal, while the state remains in the vacuum state. Alternatively, one can introduce an equivalent state formalism, where  the state  evolves along the crystal \cite{gatti2012}. By applying the equivalent of the transformation \eqref{inout} to the state, one obtains the well-known form of the  state
generated by PDC
\be
|\phi_{\rm PDC}  \rangle = |0\rangle  + \frac{1} {2} \int d\w_1 \int d\w_2 \, \psitilde (\w_1, \w_2) 
\A^\dagger (\w_1) \A^\dagger (\w_2) |0\rangle 
\label{state1}
\ee
with the biphoton amplitude $\psitilde$ being given by \eqref{psitilde} and $  |0\rangle  $ being the vacuum state. 
%%%%%%%%%%%%%%%%%%%%%%%%%%%%%%%%%%%%%%%%%%%%%%%%%%%%%%%%%%%%%%%%%%%%%%%
\section{An intuitive  picture of the spatio-temporal entanglement }
\begin{figure}[h]
\centerline{\includegraphics[width=10cm,keepaspectratio=true]{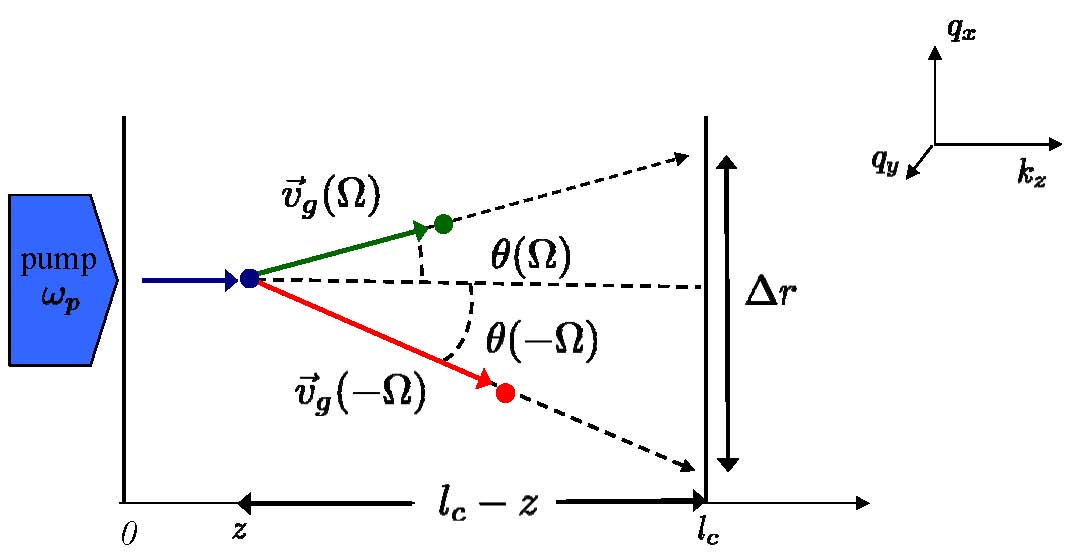}} 
\caption{\label{fig:deltax_deltatau}(Color online) Schematic representation of twin-photon propagation inside the nonlinear crystal.}% We remind that the phase matching conditions impose $\vec{q}\,(-\Om)=-\vec{q}\,(\Om)$, and therefore the twin photons propagation directions are separated by $180\degr$  in the $x\textrm{-}y$ plane.}
\end{figure}

The phase matching mechanism (the conservation of energy and momentum)  imposes a precise relationship between the frequencies of the emitted signal-idler waves and their tranverse wave vectors, which can be seen as a balance between group velocity dispersion and diffraction \cite{caspani2010,jedr2012a}. In this section we shall provide an intuitive picture of how this relationship turns into a relationship between the temporal and spatial separations of twin photons, expressed,  in the case of collinear phase matching,  by the X-structure of the spatio-temporal correlation.\\
Let us consider the limit of a plane-wave and monochromatic pump, where the only allowed processes are those where twin photons are generated in the symmetric modes $\w$, $-\w$. 
Let us focus on phase matched modes, which  satisfy the condition \footnote{For definiteness, we consider here type I e-oo phase matching, but several generalization of this formulation are possible}
\be
\Depw(\w):= \Delta(\w,-\w) = k_{sz}(\w) + k_{sz} (-\w) -k_p = 0 \, ,
\ee
that is
\begin{equation} 
\sqrt{k_{s}^{2}(\Om)-q^2}+\sqrt{k_{s}^{2}(-\Om)-q^{2}}-k_{p}=0 \; ,
\label{pmzero}
\end{equation} 
where and $k_s (\Om)$ is the signal wave-number at frequency $\omega_p/2+ \Om$. 
Equation \eqref{pmzero} defines in the 3D-space $(\q, \Om)$ a surface, identified by the phase matching curve $|q|=q_{\rm pm} (\Om)$. \\
The longitudinal momentum conservation is complemented by the transverse momentum conservation, which requires that twin photons at conjugate frequencies $\pm \Om$, are emitted with opposite transverse wave vectors: $\vec{q} (-\Om) = -\vec{q}(\Om)$, with $ |\vec{q} (\Om)|=q_{ \rm pm} (\Om)$. \\
The total derivative of Eq.\eqref{pmzero} with respect to $\Om$ (i.e. the derivative performed following the phase matching curve)  vanishes identically, and therefore:
\be
\frac{d}{d\Om} \Depw(q,\Om)  =0 \; \longrightarrow\;
\frac{\partial} {\partial \Om} \Depw(q,\Om)  = -  \frac{ \partial} {\partial q} \Depw(q,\Om)  \, q_{\rm pm}'(\Om)   \, .
\label{partial} 
\ee
Let us now assume that a pair of photons is created at a point $z$ along the crystal (see Fig.~\ref{fig:deltax_deltatau}), and let us multiply both sides of Eq.\eqref{partial} by the longitudinal distance $l_c -z$ travelled until the crystal end face. By performing the simple derivatives involved in Eq.\eqref{partial}, we obtain the identity
\be
\label{eq:der_deltapw_solo_om}
	\left[ \frac{(l_c -z) }{v_g(\Om)\cos\theta(\Om)}-\frac{(l_c -z) }{v_g(-\Om)\cos\theta(-\Om)}\right] = 
(l_c -z)  \Big[ \tan\theta(\Om) + \tan\theta(-\Om)\Big] \,  q_{\rm pm}'(\Om)  \, ,
\ee
where $v_g(\Om)=(dk_{s}(\Om) /d\Om )^{-1}$ is the group velocity at frequency $\Om$, and  $\tan[\theta(\Om)]=q_{\rm pm}(\Om)/k_{sz}[q_{\rm pm} (\Om), \Om]$.
%%%%%%%%%%%%%%%%%%%%%%%%%%%%%%%%%%%%%%%%%%
%%%%%%%%%%%%%%%%%%%%%%%%%
\par
So far, we have performed only mathematical manipulations on the phase-matching relation. Let us now interpret the result \eqref{eq:der_deltapw_solo_om}. We  picture twin photons as two wave-packets centered around the conjugate frequencies $ \pm \Om$, and propagating along directions imposed by the phase matching conditions. That is, if a  signal photon is emitted at frequency $\Om$, it will propagate at an angle $\theta(\Om)$ with respect to the $z$ axis given by $\sin[\theta(\Om)]=q_{\rm pm}(\Om)/k_{s}(\Om)$, with a velocity $v_g(\Om)$. Its twin photon will be  emitted at frequency $-\Om$ and  will propagate at an angle $\theta(-\Om)$ with a velocity $v_g(-\Om)$. 
Therefore $(l_c-z) /[v_g(\pm \Om) \cos\theta(\pm \Om)]$ represents the time employed by a signal (idler) photon to reach the crystal exit face from the point $z$ where it was created, and we can interpret the left hand side of Eq. \eqref{eq:der_deltapw_solo_om} as the temporal delay $\Delta t(\Om, z)$ acquired during propagation along the crystal by a pair of twin photon created at point $z$ with frequencies $\pm \Om$.\\
Similarly,  $(l_c-z) \tan\theta(\pm \Om)$ is the radial distance travelled by a signal (idler) photon from the point where it was created to the crystal exit face. Therefore the quantity $(l_c-z) [\tan\theta(\Om) +  \tan\theta(-\Om)] $ at the right hand side is the tranverse separation $\Delta r (\Omega, z)$  at the exit face of the crystal of a pair of twin photons created at point z with frequencies $\pm \Om$ (we remind that twin photons always travel in transversally opposite directions, therefore the sum of their radial distances from the generation point represents their transverse separation, see Fig.~\ref{fig:deltax_deltatau}).
\par
We have thus found that the phase matching condition, through Eq. \eqref{eq:der_deltapw_solo_om},  imposes a relation between the temporal delays and the transverse separations of twin photons at the crystal exit face 
\be\label{eq:tau_xi_general}
	\Delta t(\Om,z) = \Delta r(\Om,z) \, q_{\rm pm}' (\Om) \,.
\ee
The nature of this relation, i.e. the shape of the spatio temporal correlation,  depends on the shape  of the phase matching curve via its derivative $ q_{\rm pm}' (\Om)$.

A closer insight can be gained by making a Taylor espansion of the phase mismatch function \eqref{pmzero} around $\q=0,\Omega=0$ : 
\be
\Depw (\q,\Om) = \Delta_0 -\frac{q^2 l_c}{k_s} + k''_s l_c \Omega^2...
\ee
with $\Delta_0= (2k_s-k_p)l_c$ being the collinear phase mismatch at degeneracy, $k_s=k_s(\Omega=0)$, and 
$k''_s= d^2 k_s(\Omega) /d\Omega^2 |_{\Omega=0}$. 
We can consider the two   limiting cases: \\

\noindent
{\bf a) Collinear phase matching $\Delta_0=0$}. In a large neighbourhood of  $\Om=0$, the phase matching curve corresponds to two straight lines with opposite slopes (see also Fig.\ref{fig2}a) 
\be
q_{\rm pm}(\Om)= \pm \sqrt{k_s k''_s} \, \Om 
\ee
(where the $+$ or $-$ sign has to be taken for $\Omega$ positive or negative, respectively), so that 
$
q_{\rm pm}' (\Om)= \pm \sqrt{k_s k''_s} $ and Eq.(\ref{eq:tau_xi_general})  becomes
\be
\label{asymptotes}
	\Delta t(\Om,z)=\pm  \sqrt{k_s k''_s} \, \Delta r (\Om,z) \,. 
\ee
This equation tells us that,  whatever was the position along the crystal where twin photons were generated, whatever is their temporal frequency, they arrive at the end face of the crystal with a temporal delay  proportional to their spatial  transverse separation (with a plus or minus sign). Since we do not know where along the crystal and with which frequency twin photons were created, there is  un uncertainty in their  arrival times and transverse separations. 
However, when they are found separated by a distance $\Delta r$, they will be also separated in time by an  amount $ \Delta t=\pm
\sqrt{k_s k''_s} \Delta r$.  
This entirely classical reasoning thus predicts a strong coupling between spatial and temporal degrees of freedom of twin photons. In particular Eq.\eqref{asymptotes}, when seen in the full 3D space $\Delta x, \Delta y, \Delta t$,  describes a biconical surface, thus predicting that the spatio-temporal correlation has a non-factorable biconical geometry, which appears as an {\em  X}  in any plane containing one spatial coordinate and time.  
\\

\noindent
{\bf b)  Non-collinear phase matching $ \Delta_0 \gg k''_s l_c \Omega^2  $}. In this conditions,   in the neighbourhood of $\Omega=0$ the phase matching curve  basically becomes a horizontal straight line (see also Fig. \ref{fig2}c) : 
\be
q_{\rm pm} (\Om) = \sqrt{k_s \Delta_0/l_c + k_s k''_s \Omega^2}  \approx   \sqrt{k_s\Delta_0/l_c}   \quad {\rm for} \; 
\Delta_0 \gg k''_s l_c \Omega^2  
\ee 
so that $q_{\rm pm}'(\Omega) \approx 0$ and Eq.(\ref{eq:tau_xi_general})  predicts that 
\be
\label{noncollinear}
	\Delta t(\Om,z)\approx 0  \, ,
\ee
independently from the radial separation $\Delta r$. This result tells us that twin photons are most likely to be found toghether in time at the exit face of the crystal, irrespectively from their position in their transverse cross section, and  suggests that for increasing values of  the parameter $\Delta_0$ the spatial and temporal degree of freedom of twin photon should become progressively independent. This feature is indeed reflected by the structure of the quantum spatio-temporal correlation  that will be   described in the next section.
%%%%%%%%%%%%%%%%
\section{Formal calculation of the spatio-temporal correlation} 
Besides the classical-looking picture developed in the previous section, a full quantum calculation of the spatio-temporal correlation function can be performed, as done in \cite{gatti2009,caspani2010}. 

\begin{figure}[h]
\centerline{\includegraphics[width=11cm,keepaspectratio=true]{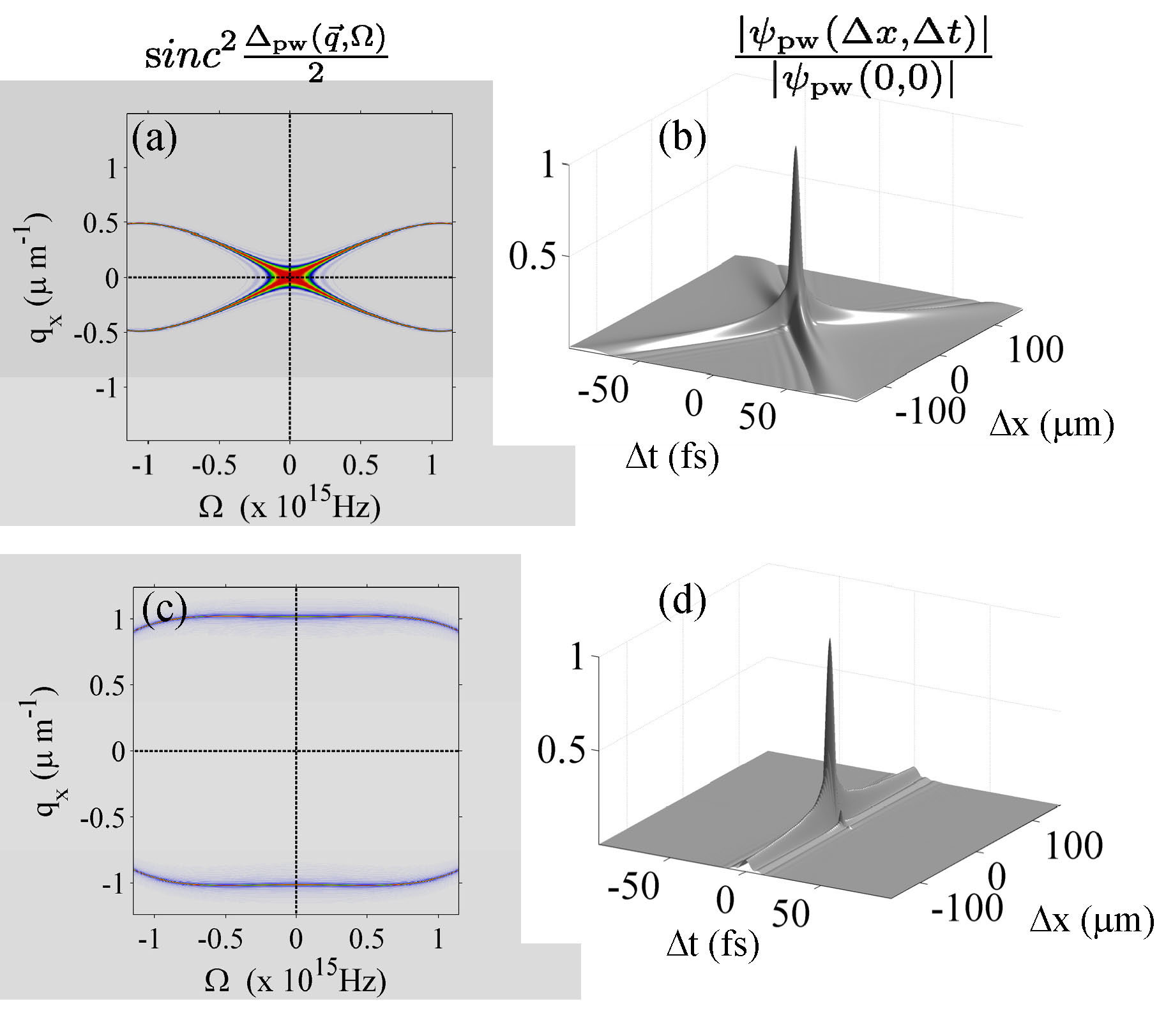}} 
\caption{(Color online) The spatio temporal correlation of twin photons calculated from Eq. \eqref{psipwb}, for a 4 mm BBO crystal pumped at 527.5 nm.  The first and second row are plotted for collinear $\Delta_0=0$ and non-collinear $\Delta_0= 419 $ phase matching, respectively (the angles between the crystal axis and the z-axis are $22.9^\circ$ and $28^\circ$, respectively) .
Panels (a) and (c) show the function $\sinc^2 [\Depw (\q,\Om)/2] $, centered around the phase matching curve.  in the plane $(q_x, \Omega)$, while panels (b) and (d) are the corresponding biphoton correlations as a function of the temporal  and spatial  separation $\Delta t$, $\Delta x$. Coupling  parameter $g=10^{-3}$ } 
\label{fig2}
\end{figure}

In the low-gain regime the spatio-temporal correlation properties of twin photons are entirely described by the biphoton amplitude in the direct domain : 
\bea
\psi (\vxi, \vxi\,^\prime) : = \langle \A(\vxi) \A(\vxi\,^\prime) \rangle 
 = \frac{1}{(2\pi)^3} \int d\w \int d\wprime \psitilde (\w, \wprime) \e^{i \vxi\cdot \w + i \vxi\,^\prime \cdot \wprime}
\label{psi}
\eea
where $\psitilde$ is given by Eq. \eqref{psitilde},  and  we remind that $\vxi= (\rr, t)$,  where $\rr$ is the transverse coordinate in the cross-section at the exit face of the crystal slab. Thus the function \eqref{psi} gives the probability amplitude of finding a pair of photons at times $t,t' $ and positions $\rr, \rr'$, respectively,  
at the end face of the crystal.\\
A rather simple expression can be written in the nearly plane-wave pump  approximation (see \cite{caspani2010} for details), where the pump spectral profile is assumed to be narrow enough (the pump pulse has a wide enough cross-section and long enough duration), so that the following approximation holds: 
\be
\psitilde(\w,\wprime) \approx 
\frac{ g} { (2 \pi)^{3/2} }   \Aptilde(\w+\wprime,l_c) \e^{+i\frac{\Delta(\w,-\w)}{2} } \sinc \frac{\Delta(\w,-\w) }{2}  \, .
\label{psitildepw}
\ee
The limit \eqref{psitildepw} is not difficult to reach, and very often corresponds to the actual experimental implementations of PDC (see e.g. the experiment in \cite{jedr2012a,jedr2012b}). As explained in \cite{caspani2010},  it  requires that the pump pulse is longer than the temporal delay between the pump and the signal wave due to group velocity mismatch (for a 4 mm BBO-Beta Barium Borate- crystal $\approx$ 350 fs), and that its cross section is larger than their spatial walk-off (for a 4 mm BBO crystal $\approx$ 220 $\mu$m).\\
In this limit, the biphoton spatio-temporal correlation takes the factorized form: 
\bea
\psi(\vxi, \vxi' ) &=& \Ap \left(\frac {\vxi+ \vxi'} {2}, l_c \right) \psi_{\rm pw} (\vxi' - \vxi) \label{psipwa}\\
\psi_{\rm pw} (\vxi' - \vxi) &=& g \int \frac{d\w} {(2\pi)^3}  
\e^{i (\vxi'-\vxi)  \cdot \w}   \sinc{\frac {\Depw (\w)}{2} } \,  \e^{i  \frac {\Depw (\w)}{2}   } 
\label{psipwb} \, .   
\eea
Thus, when plotted as a function of the mean spatio-temporal coordinate of twin photons,   the  biphotonic correlation  follows the slow variation of the pump spatio-temporal profile. On the other hand,  as a function of the temporal and spatial separation of twin photons, it has a fast variation represented  by  the Fourier transform of the $\sinc$ function, strongly peaked around the phase matching curve $\Depw=0$. Fig.\ref{fig2} plots  examples of such a spatio-temporal correlation, calculated in the two limiting cases of exactly collinear and extremely non-collinear phase matching discussed in the former section.
%%%%%%%%%%%%%%%%%%%%%%%%%%%%%%%%%%%%%%%%%%
%%%%%%%%%%%%%%%%%%%%%%%%%
This figure shows  a clear agreement between  the fully quantum mechanical calculation and  the qualitative prediction based on the classical argument of the former section. In fact, for collinear phase matching the quantum correlation of Fig.\ref{fig2}b  exhibits a X-shaped geometry, which would appear as a bicone in the 3D space, implying a proportionality between the temporal delays of twin photons and their spatial separation. As discussed in detail in \cite{gatti2009,caspani2010} and demonstrated experimentally in \cite{jedr2012b} the asymptotes of the X-structure lie exactly on the lines $\Delta t = \pm \sqrt{k_s k_s''} \Delta r$ as predicted by the classical argument  of Eq.\eqref{asymptotes}. Conversely, for non-collinear phase matching, the spatio-temporal quantum correlation progressively shrinks and tends to assume the sigar-like shape of Fig.\ref{fig2}d, where the temporal and spatial separation of twin photon are basically independent. 
%%%%%
\section{Schmidt number of the spatio-temporal entanglement}
\begin{figure}[h]
\centerline{\includegraphics[width=12cm,keepaspectratio=true]{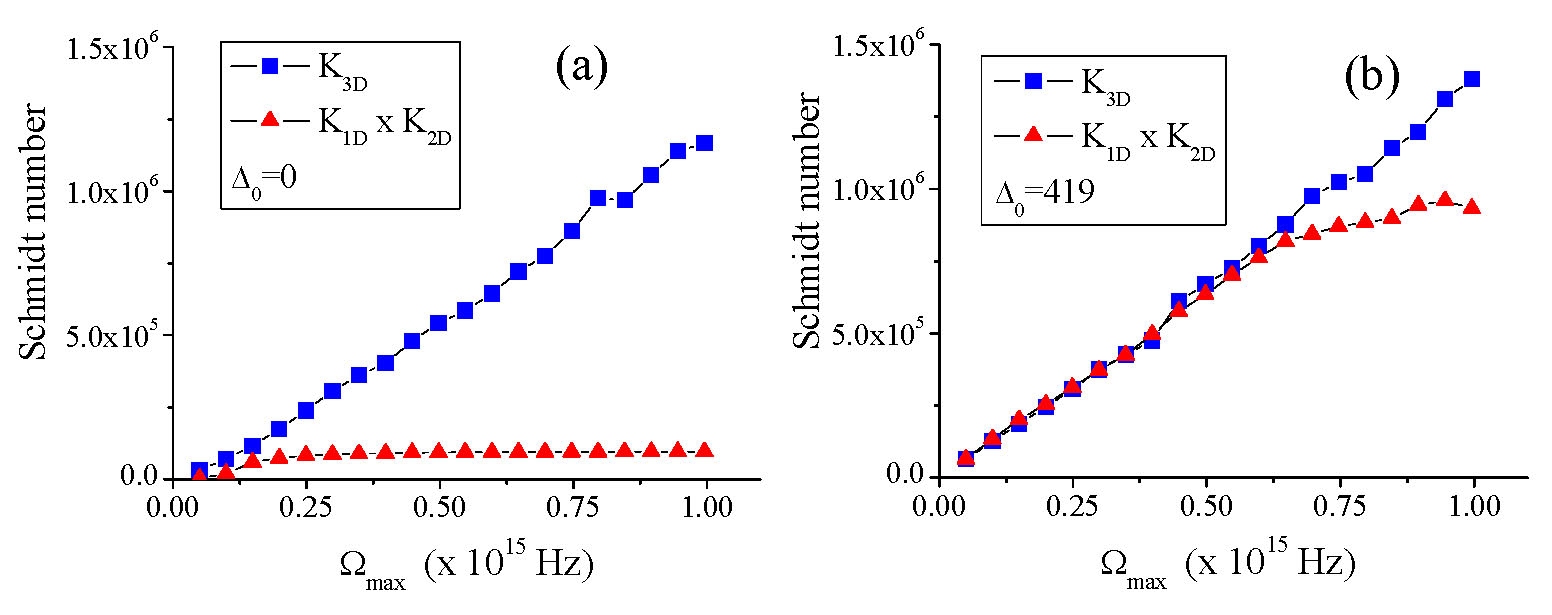}} 
\caption{(Color online) Comparison between the Schmidt number calculated in the full 3D spatio-temporal model ($K_{3D}$),   and the product of Schmidt numbers calculated in a purely spatial ($K_{2D}$) and temporal model ($K_{1D}$) for PDC, as a function of the detected temporal bandwidth. (a) Collinear $\Delta_0=0$, (b) non-collinear $\Delta_0= 419 $ phase matching. Evident is the transition from a state non-factorable in space and time towards a more factorable state.  $q_{max}=1.2 \mu{\rm m}^{-1}$,  pump waist $600 \mu$m, pump duration 1 ps.  Other parameters as in Fig.\ref{fig2}} 
\label{fig3}
\end{figure}

In the framework of the high-dimensional entanglement realized by spontaneous PDC, an important question corcerns  the number of entangled modes generated by the process or in other words of the degree of entanglementof the state. This can by quantified by the Schmidt number \cite{ekert1994,parker2000}. Traditional evaluation of the Schmidt dimensionality of parametric-down conversion typically have concentrated on either the purely temporal entanglement \cite{law2000,mikhailova2008,patera2010,avenhaus2009}  or on the purely transverse spatial entanglement \cite{law2004,exter2006,dilorenzo2009,dilorenzo2010}. In a recent work \cite{gatti2012} some of us studied the Schmidt dimensionality of the two-photon state generated by PDC in the framework of a fully 3D spatio-temporal model, and showed that in conditions of collinear phase matching this cannot be reduced to the product of Schmidt number calculated in models restricted to the purely temporal and purely spatial domain.\\
 This is basically the result shown in Fig.\ref{fig3}a, where the Schmidt number calculated in the 3D spatio-temporal model for PDC ($K_{3D}$) in collinear conditions is compared with the product of the temporal ($K_{1D}$)  Schmidt number times the spatial ($K_{2D}$) Schmidt number, as a function of the collected temporal bandwidth $\Omega_{max}$. In this plot we see that,  as soon as the collected bandwidth exceeds $\approx 10^{14} Hz$,  the degree of the spatio-temporal entanglement does not factorizes into its purely temporal and spatial counterparts, which is again an evidence of the space-time coupling occurring in conditions of collinear phase matching, predicted in Sec. 3 with qualitative arguments and evidenced in Sec.4 by the non-factorable X-correlation of Fig\ref{fig2}b .\\ 
On the other side, the panel (b) of Fig.\ref{fig3} shows the same comparison in conditions of non-collinear phase matching. In this case $K_{3D} \approx K_{1D} \times K_{22013
D}  $ over a much larger range of collected spectral bandwidths, 
%some difference being appreciable only for $\Omega_{max} > 0.5 10^{15} $Hz.  
Thus the analisys of the Schmidt number of the entanglement gives again evidence, from a completely different perspective, of the transition from a state where the spatial and temporal degrees of freedom are strongly coupled to a state whose spatial and temporal properties can be considered almost independent. In this transition, the classical phase matching conditions play a major role. 

%%%%%%%%%%%%%%%%%%%%%%%%%%%%%%%%%%%%%%%%%%
%%%%%%%%%%%%%%%%%%%%%%%%%
The calculation of the Schmidt numbers reported in Fig.\ref{fig3} proceeds along the following steps (see \cite{gatti2012} for more details):\\
First of all,  a simple bipartition of the system is introduced, by considering the action of a 50:50 beam splitter on the downconverted light at the output of the crystal slab. Second,  the two-photon state conditioned to the joint detection of a pair of photons in the  output modes $\A_1, \A_2$ of the beam-splitter is considered: 
\be
| \phi \rangle = \int d\w_1 \int d\w_2 \, \psitilde (\w_1, \w_2) 
\A_1^\dagger (\w_1) \A_2^\dagger (\w_2) |0\rangle_1 |0\rangle_2 
\label{state2}
\, .
\ee
Finally the Schmidt number is introduced as 
\be
K= \frac{1}{ {\mathrm Tr} \{\rho_1^2\} }
\label{kappa}
\ee
where $\rho_1$ is the reduced density matrix of the subsystem 1
\be
\rho_1 = {\mathrm Tr}_2 \{ \frac{|\phi \rangle \, \langle \phi | } {\langle \phi| \phi\rangle}         \} \, .
\ee
As reported in detail in \cite{gatti2012}, starting from the espression \eqref{state2} of the conditional two-photon state state, an integral formula for the Schmidt number can be derived, which reads: 
\begin{subequations}
\label{formula}
\be
K = \frac{N^2} {B} 
\label{kintegral}
\ee
where
\bea
N &= \int d\w_1  \langle \phi_{\rm PDC} |   \A^{\dagger} (\w_1)   \A (\w_1) | \phi_{\rm PDC} \rangle 
= \int d\w_1 \int d\w_2 \left| \psitilde(\w_1, \w_2) \right|^2
\label{N}\\
B &= \int d\w_1 \int d\w_2 \int d\w_3 \int d\w_4  \psitilde (\w_1,\w_2) \psitilde(\w_3,\w_4) 
 \psitilde^* (\w_1,\w_4) \psitilde^* (\w_3,\w_2) 
\label{B} 
\eea
\end{subequations} 
Notice that by using simple mathematical properties of the Fourier transform, relations \eqref{formula} can be turned into identical relations involving the biphoton amplitude $\psi (\vxi, \vxi\,^\prime)$ in the {\em direct domain} in place of 
the amplitude  $\psitilde (\w,\wprime)$ in the Fourier domain (the two amplitudes are in fact linked by a Fourier transform, see Eq\eqref{psi}): the degree of entanglement is independent  wether one looks at photons generated in the Fourier space or in the conjugate one.\\
Ref. \cite{gatti2012} described various approximations for formula \eqref{formula}, which lead to interesting analytical or semianalytical results for the Schmidt number. For simplicity, our Fig.\ref{fig3} presents only the numerical results for $K$, where the integrals in Eqs.(\ref{N}, \ref{B}) are evaluated with a Montecarlo method, and $\psitilde$ is given by Eq. \eqref{psitilde}.
In the 3D case,  each of the integrals in the above formulas is evaluated in the 3D space $\w_i=(\q_i,\Omega_i)$, with $|\q_i| $ and $\Om_i$ ranging from zero up to  some maximum values $q_{max}$ and $\Om_{max}$, which  simulate the presence of spatial or temporal filters (in Fig.\ref{fig3}, $q_{max}$ is fixed to a value large enough to include all the phase matching curve).
In the 2D model,  the integration variable becomes the 2D spatial coordinate $\w_i \to \q_i$, ranging up to $q_{max}$, 
while the temporal frequency is fixed at $\Om=0$. In the 1D model, the integration variable is only the temporal frequency  $\w_i \to \Om_i \; \epsilon\; (0,\Om_{max})$, , while the spatial coordinate is fixed to $|\q| = q_{\rm pm} (\Omega=0)$. Precisely in panel (a) $|\q| = 0$, while in panel (b) $|\q| = \sqrt{k_s k_s''/ l_c} $. 
%%%%%%%%%%%%%%%%%%%%%%%%%%%%
\section{Conclusions} 
In this work we have developed an intuitive picture of the spatial and temporal properties of the entangled state generated by parametric-down-conversion, where twin photons are described as classical wave-packets. In this way we have shown that the phase matching mechanism, which links the angle and frequency of emission of light, can be translated into a precise relation between the temporal and transverse spatial separation of twin photons when they emerge from the nonlinear crystal.
Althought our reasoning is  tailored to describe type I PDC, we believe that it is quite general, and could be easily adapted to describe other nonlinear processes governed by phase matching. 
\par
This coupling between spatial and temporal degrees of freedom, described by classical arguments, turns out to determine in a precise way the spatio-temporal structure of the quantum entanglement. In particular, the quantum  biphotonic correlation is shown to be skewed in  space and time along the same trajectories predicted by the classical argument.  The space-time coupling in the parametric process manifests itself also in 
the properties of the Schmidt number, which, in this case, gives an estimate of the number of modes effectively involved in the entanglement: in general it turns out that the number of spatio-temporal modes does not equal to the product of purely spatial and purely temporal modes.
\par
Finally, we have shown that by properly manipulating the phase matching conditions it is possible to modify the relation between temporal and spatial separation of twin photons, thereby modifying  the structure of the entangled state. In particular by changing the crystal
tuning angle so to achieve strongly non collinear phase matching, we see a transition from
a non-separable X-geometry to an almost separable geometry in space and time. This latter
situation could be indeed very interesting for applications because twin photons can be relatively localized in time within few fs without the need of resolving their
positions, i.e. an extremely broad temporal frequency bandwidth can be achieved independently from the spatial measurement.
\vspace*{-6pt}   %ONLY NECESSARY
\section*{Acknowledgments}
This work was realized in the framework of the Fet Open project of EC 221906 HIDEAS.
\vspace*{-6pt}   %ONLY NECESSARY

\end{document}